\documentclass[reprint,superscriptaddress,amsmath,amssymb,
aps,prl,floatfix]{revtex4-1}

\usepackage{bm}
\usepackage{float}
\usepackage{graphicx}
\usepackage[usenames,dvipsnames]{color}
\usepackage[normalem]{ulem}
\usepackage[svgnames]{xcolor}
\usepackage{bm}
\usepackage{multirow}
\usepackage{float}
\usepackage{titlesec}
\usepackage[utf8]{inputenc}

\begin{document}
\preprint{APS/123-QED}


\title{Quasi-isotropic orbital magnetoresistance in lightly doped SrTiO$_{3}$}

\author{ Cl\'ement Collignon}
\altaffiliation[Present address: ]{Department of Physics, Massachusetts Institute of Technology, Cambridge, Massachusetts
02139, USA}
\affiliation{JEIP,  USR 3573 CNRS, Coll\`ege de France, PSL Research University, 11, place Marcelin Berthelot, 75231 Paris Cedex 05, France}
\affiliation{Laboratoire de Physique et d'Étude des Matériaux (ESPCI Paris - CNRS - Sorbonne Universit\'e), PSL Research University, 75005 Paris, France}

\author{Yudai Awashima}
\affiliation{Department of Engineering Science, University of Electro-Communications, Chofu, Tokyo 182-8585,
Japan}

\author{Ravi}
\affiliation{Laboratoire de Physique et d'Étude des Matériaux (ESPCI Paris - CNRS - Sorbonne Universit\'e), PSL Research University, 75005 Paris, France}

\author{Xiao Lin}
\altaffiliation[Present address: ]{Westlake University, 310024 Hangzhou, China }
\affiliation{Laboratoire de Physique et d'Étude des Matériaux (ESPCI Paris - CNRS - Sorbonne Universit\'e), PSL Research University, 75005 Paris, France}

\author{Carl Willem Rischau}
\altaffiliation[Present address: ]{Department of Quantum Matter Physics,  University of Geneva, 24 Quai  Ernest-Ansermet,  1211 Geneva 4,  Switzerland}
\affiliation{Laboratoire de Physique et d'Étude des Matériaux (ESPCI Paris - CNRS - Sorbonne Universit\'e), PSL Research University, 75005 Paris, France}

\author{Anissa Acheche}
\affiliation{JEIP,  USR 3573 CNRS, Coll\`ege de France, PSL Research University, 11, place Marcelin Berthelot, 75231 Paris Cedex 05, France}

\author{Baptiste Vignolle}
\affiliation{Laboratoire National des Champs  Magn\'etiques Intenses (LNCMI-EMFL), CNRS ,UGA, UPS, INSA, Grenoble/Toulouse, France}
\affiliation{Institut de Chimie de la Matière Condensée, Bordeaux, France}

\author{Cyril Proust}
\affiliation{Laboratoire National des Champs  Magn\'etiques Intenses (LNCMI-EMFL), CNRS ,UGA, UPS, INSA, Grenoble/Toulouse, France}

\author{Yuki Fuseya}
\affiliation{Department of Engineering Science, University of Electro-Communications, Chofu, Tokyo 182-8585,
Japan}
\affiliation{Institute for Advanced Science, University of Electro-Communications, Chofu, Tokyo 182-8585, Japan}

\author{Kamran Behnia}
\affiliation{Laboratoire de Physique et d'Étude des Matériaux (ESPCI Paris - CNRS - Sorbonne Universit\'e), PSL Research University, 75005 Paris, France}

\author{Benoit Fauqu\'e}%
 \email{benoit.fauque@espci.fr}
\affiliation{JEIP,  USR 3573 CNRS, Coll\`ege de France, PSL Research University, 11, place Marcelin Berthelot, 75231 Paris Cedex 05, France}

\date{\today}

\begin{abstract}
A magnetic field parallel to an electrical current does not produce a Lorentz force on the charge carriers. Therefore, orbital longitudinal magnetoresistance is unexpected. Here we report on the observation of a large and non saturating magnetoresistance in lightly doped SrTiO$_{3-x}$ independent of the relative orientation of current and magnetic field. We show that this quasi-isotropic magnetoresistance can be explained if the carrier mobility along all orientations smoothly decreases with magnetic field. This anomalous regime is restricted to low concentrations when the dipolar correlation length is longer than the distance between carriers. We identify cyclotron motion of electrons in a potential landscape tailored by polar domains as the cradle of quasi-isotropic orbital magnetoresistance. The result emerges as a challenge to theory and may be a generic feature of lightly-doped quantum paralectric materials.    

\end{abstract}

\maketitle

Magnetoresistance (MR), the change in electrical resistivity  under the application of a magnetic field is an old topic in condensed matter physics~\cite{Pippard89}. It can be simply understood as a consequence of the Lorentz force exerted on mobile electrons by the magnetic field. This orbital magnetoresistance (which neglects the spin of electrons) is largest when the magnetic field is perpendicular to the  electrical current. The transverse magnetoresistance (labelled TMR) is expected to increase quadratically with magnetic field at low fields and then saturate at high fields.  The boundary between the two regimes is set by $\mu_{H}B\approx1$ (where $\mu_{H}$ is the Hall mobility). When the field and the current are parallel, we are in presence of longitudinal magnetoresistance (labelled LMR), expected to be negligibly small due to the cancellation of the Lorentz force. 

However, this simple picture is known to fail in numerous cases. Non-saturating linear TMR has been observed in electronic systems ranging from potassium \cite{Pippard89}, to doped silicon \cite{Delmo2009}, 2D electron gas system \cite{Khouri2016}, 3D doped semi-conductors and Dirac materials \cite{Xu1997,Hu2005,Kozlova2012,Schneider2014,Fauque2013,Novak15,Narayanan2015,Xiong_2016}, density wave materials \cite{Feng2019}, or correlated materials \cite{Hayes2016}. LMR, one order of magnitude smaller (than and with an opposite sign to) its transverse counterpart has been observed in silver chalcogenides \cite{Hu2005} and topological materials \cite{Wiedmann2016,Reis_2016}. The exact conditions for the emergence of a sizeable LMR is the subject of  ongoing debate \cite{Pal2010,Burkov2015,Goswami2015}.

\begin{figure}
\begin{center}
\centering
\includegraphics[width=0.5\textwidth]{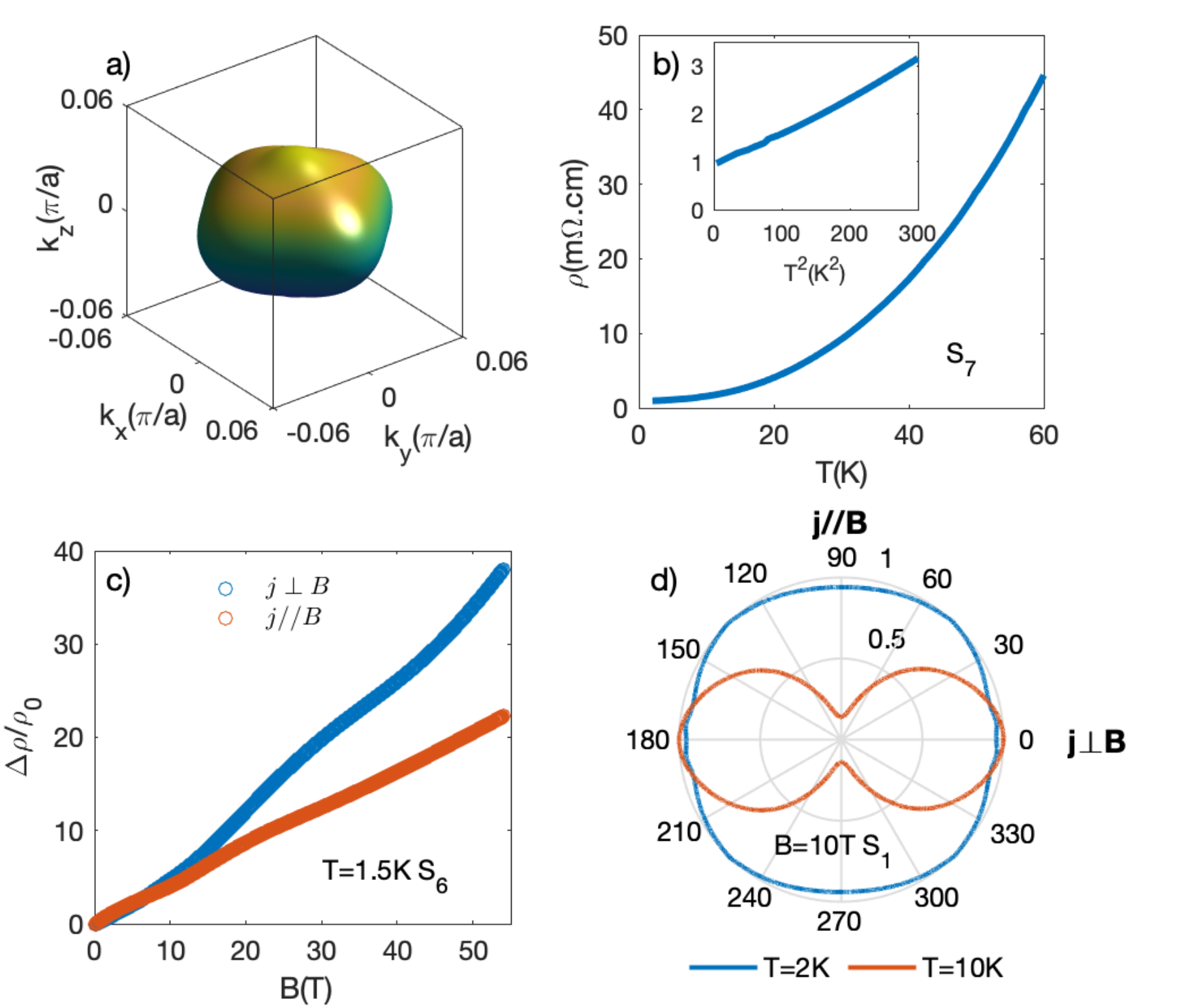}
\caption{ {\bf{Electrical transport properties of a lightly doped SrTiO$_{3-x}$ at low temperature} :} a) Fermi surface of the lower band of SrTiO$_{3}$ according to \cite{Allen2013} b) Temperature dependence of the resistivity ($\rho$) of sample S$_{7}$. Insert: $\rho$ {\it{vs}} $T^2$. c) Transverse ($j \perp B$) and longitudinal ($j // B$) magnetoresistance ($\frac{\Delta \rho}{\rho_0}$=$\frac{\rho(B)-\rho(B=0)}{\rho(B=0)}$) as function of the magnetic field up to 54 T, at $T=1.5$ K, for S$_{6}$ ($n_H$(S$_{6})=3.2\times 10^{17}$ cm$^{-3}$). d) Normalised angular magneto-resistance of S$_{1}$ ($n_H($S$_1)=6.5\times 10^{16}$ cm$^{-3}$) in polar plot at $B=10$ T for two temperatures :  $T=2$ K (in blue) and $T=10$ K (in red). $\theta=0^\circ$ and $90^\circ$ correspond respectively to ${\bf{j}}\perp {\bf{B}}$ and ${\bf{j}}//{\bf{B}}$)}
\label{Fig1}
\end{center}
\end{figure}

In this paper, we report on the case of lightly doped SrTiO$_{3}$. Undoped strontium titanate is an incipient ferroelectric, dubbed quantum paraelectric \cite{Muller1989}, which can be turned into a metal by non-covalent substitution or by removing oxygen \cite{Spinelli2010}. This dilute metal \cite{Collignon2019} has attracted recent attention \cite{Zhou2019,kumar2020} due to the persistence of $T$-square resistivity in absence of Umklapp and interband scattering among electrons \cite{Lin2015sc} and the unexpected survival of metallicity at high temperatures \cite{Collignon2020}. We will see below that its magnetoresistance is also remarkably non-trivial. In contrast with any other documented material, it shows a large and quasi-linear TMR, accompanied by a positive LMR of comparable amplitude. Intriguingly, the amplitude of magnetoresistance depends only on the amplitude of the magnetic field, independent of the mutual orientation of the current and the magnetic field.  We will show that this unusual quasi-isotropic magnetoresistance is restricted to a range of  doping where the inter-electron distance exceeds the typical size of a polar domain. The observation implies that this phenomenon is driven by the interplay between cyclotron orbits and the potential landscape shaped by polar domains and suggests that it may be generic to lightly doped quantum paraelectrics. 

Fig.\ref{Fig1} presents our main result. When the carrier density in  SrTiO$_{3-\delta}$ is $n_{H}=3\times10^{17}$ cm$^{-3}$, there is a single Fermi pocket at the center of the Brillouin zone shown Fig.\ref{Fig1}a). This is what is expected by band calculations \cite{marel2011} and found by quantum oscillations \cite{Uwe_1985,Allen2013,Lin2014}. Nevertheless, not only this dilute metal displays a $T^{2}$behavior (see Fig.\ref{Fig1}b)), but it also responds to magnetic field in a striking manner. Upon the application of a magnetic field of 54 T, there is a forty (twenty)-fold enhancement of resistance for the transverse (longitudinal) configuration (see Fig.\ref{Fig1}c)). In both cases, the evolution with field is quasi-linear and there is no sign of saturation even if the high field regime ($\mu_{H}B>>1$) is clearly attained.

\begin{figure*}
\centering
\makebox{\includegraphics[width=1.0\textwidth]{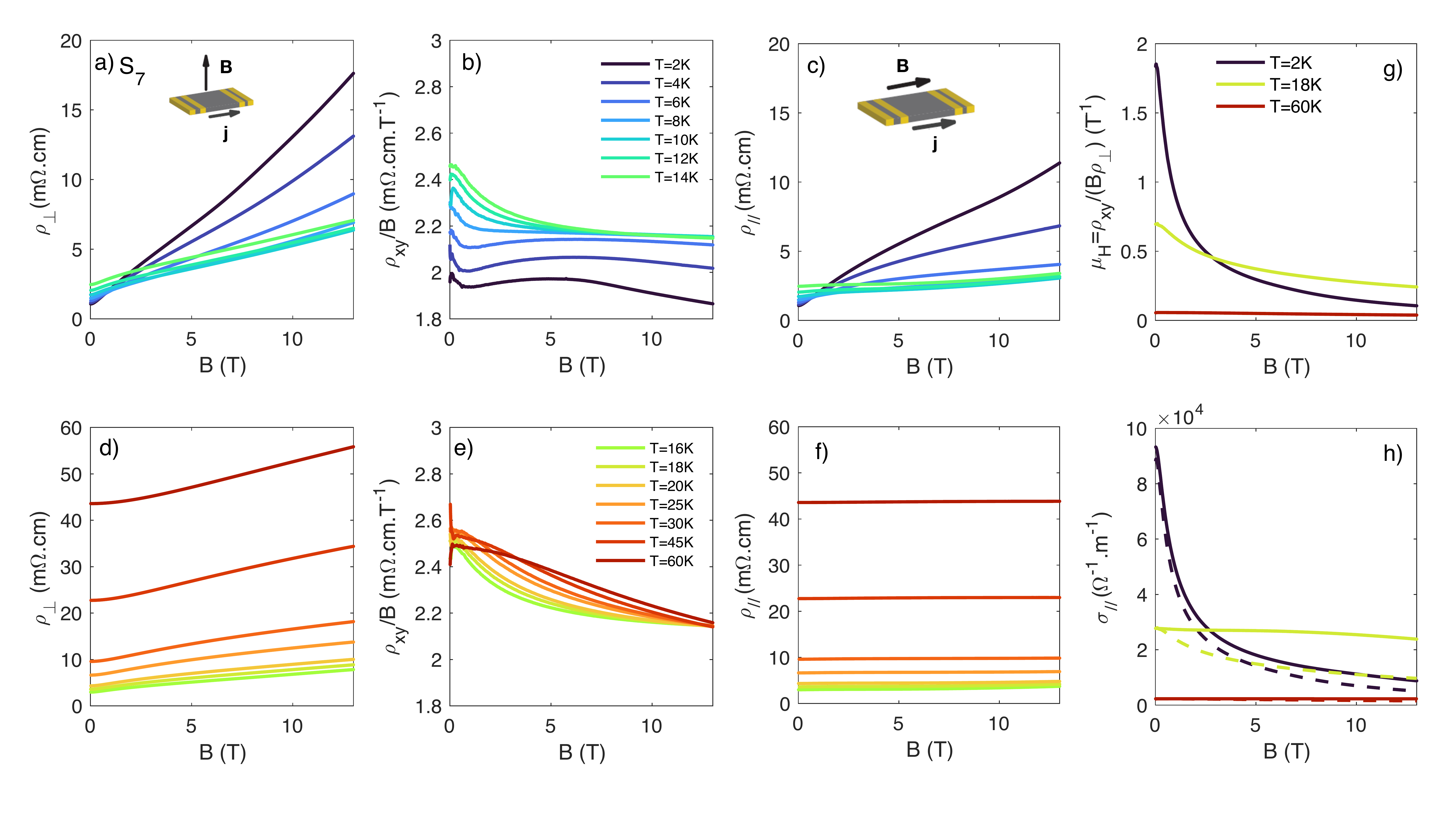}}
\caption{{\bf{Temperature dependence of the transverse and longitudinal magneto-resistance}} : a)-c) Field dependence of the resistivity for ${\bf{j}}\perp{\bf{B}}$ ($\rho_{\perp}$), $\frac{\rho_{xy}}{B}$ and the resistivity for ${\bf{j}}//{\bf{B}}$ ($\rho_{//}$) from $T=2$ to 14 K for S$_7$ ($n_H($S$_7)=3.3\times 10^{17}$ cm$^{-3}$) d)-f) same as a)-c) from $T=16$ to 60 K. g) Field dependence of the Hall mobility $\mu_H=\frac{\rho_{xy}}{B\rho_{\perp}}$ for $T=2$, 18 and 60 K. h) Longitudinal magnetoconductivity ($\frac{1}{\rho_{//}}$) compared with $\sigma_{//}=ne\mu_H(B)$ with the deduced $\mu_{H}(B)$ shown on g).}.
\label{FigTemp}
\end{figure*}
A polar plot of the normalised angular magnetoresistance (AMR) at a fixed magnetic field for another sample (S$_{1}$) with a slightly lower carrier density ($n_{H}($S$_{1})=6.5 \times10^{16}$ cm$^{-3}$) is shown on Fig.\ref{Fig1}d). The magnetic field rotates from the transverse ($\theta=0^\circ$) to the longitudinal ($\theta=90^\circ$) configuration at two different temperatures. While at $T=10$ K, the longitudinal magnetoresistance shrinks towards zero, at $T=2$ K  the magnetoresistance is quasi-isotropic and the relative direction of the magnetic field and the current injection barely affects its amplitude. 

When a magnetic field is aligned along the $z$-axis, in presence of a single-component Fermi surface, the three  components of the conductivity tensor have remarkably simple expressions: 

\begin{equation}
\sigma_{zz}=\sigma_{//}=ne \mu_H
\end{equation}

\begin{equation}
\sigma_{xx}=\sigma_{\perp}=\frac{ne \mu_H}{1+\mu^2_{H}B^2}
\end{equation}

\begin{equation}
\sigma_{xy}=\mu_{H}B\frac{ne\mu_H }{1+\mu^2_{H}B^2}
\end{equation}

Now, if $\mu_H$ remains constant as a function of magnetic field, one does not expect the longitudinal magnetoresistance, since  $\rho_{//}=\sigma_{//}^{-1}$ would not depend on magnetic field. One would not even expect a transverse magnetoresistance, because the same is true for $\rho_{\perp}=\frac {\sigma_{\perp}}{\sigma_{\perp}^2+\sigma_{xy}^2}= \frac {1}{ne \mu_H}$. 

These equations hold in presence of a quadratic dispersion when the effective mass $m^{*}$ and the Hall mobility, $\mu_H= e\tau /m^{*}$ are well defined. The Fermi pocket associated with the lowest band in dilute metallic strontium titanate is not an ellipsoid. This can lead to a finite TMR and LMR \cite{Pippard89,Pal2010}. However, as discussed in the supplement \cite{SM}, the  results computed using the specific geometry of the Fermi surface are well below the experimentally observed magnitudes at low temperature. As we will see below, to explain our result, one needs to assume a field-dependent $\mu_H$.

\begin{figure}
\begin{center}
\includegraphics[angle=0,width=8.5cm]{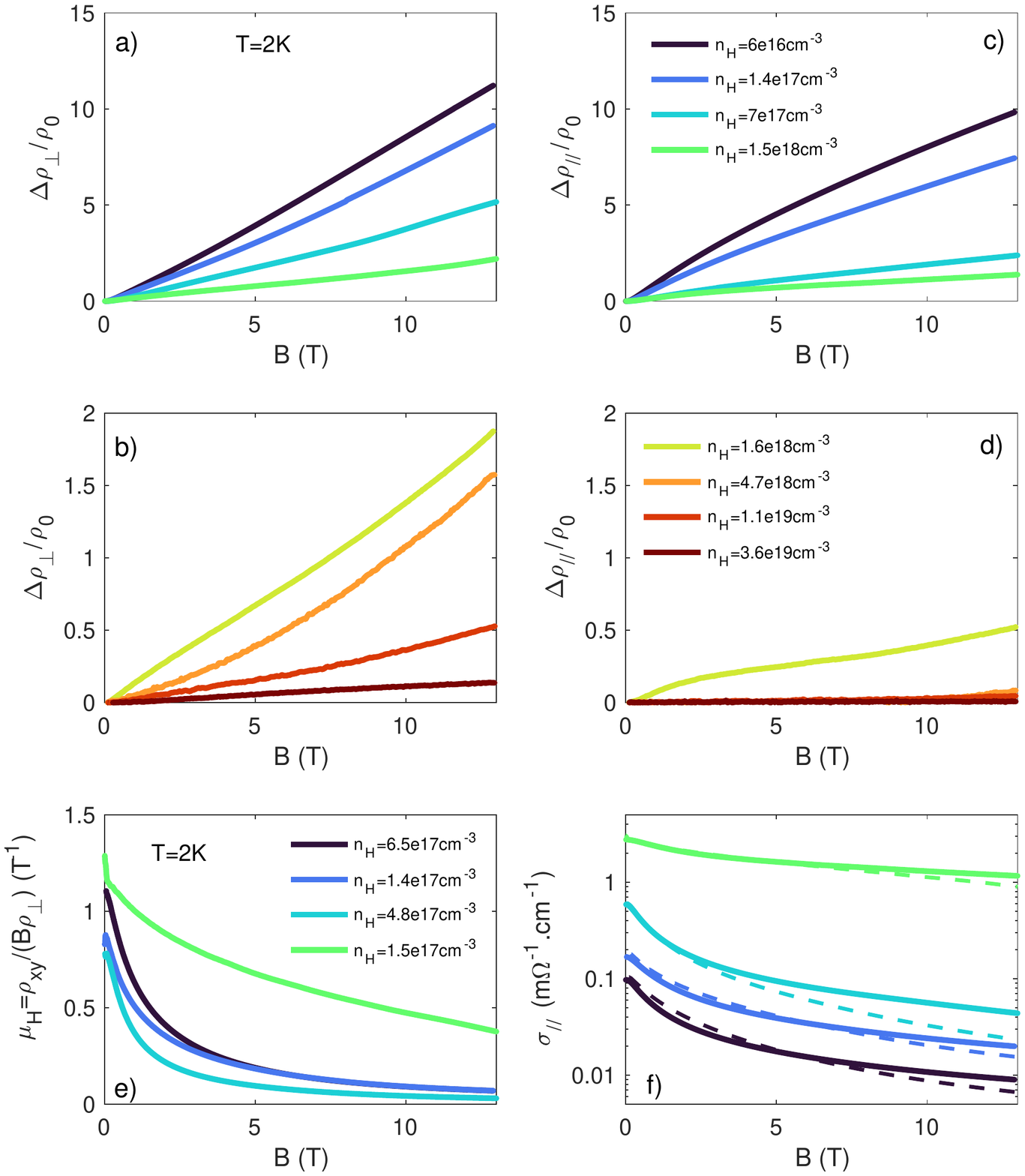}
\caption{ {\bf{Doping dependence of the transverse and the longitudinal magnetoresistance at $T=2$ K}} : a) and b) TMR for n$_H$ ranging from $6.5\times 10^{16}$ to $3.6\times 10^{19}$ cm$^{-3}$. c) and d) same as a) and b) for the LMR. e) Field dependence of Hall mobility ($\mu_{H}$) deduced from the TMR and the Hall effect measurements in four low doped samples. f) Comparison of the longitudinal magnetoconductance ($\frac{1}{\rho_{//}}$) with $\sigma_{//}=ne\mu_H(B)$ with the deduced $\mu_{H}(B)$ shown on e).}
\label{FigDop}
\end{center}
\end{figure}

Fig.\ref{FigTemp} shows the evolution of the quasi-isotropic magnetoresistance with temperature. The amplitude of the TMR decreases with warming (see Fig.\ref{FigTemp}a) and d)). The same is true for the LMR (see Fig.\ref{FigTemp}c) and f)). On the other hand, the Hall coefficient is barely temperature-dependent (see Fig.\ref{FigTemp}b) and e)). Upon warming, the longitudinal magnetoresistance decreases faster than its transverse counterpart and above 14 K it almost vanishes (Fig.\ref{FigTemp}c) and d)). Above this temperature, a small TMR persists with an amplitude comparable with what the semi-classical theory expects (see supplement \cite{SM}). Fig.\ref{FigDop} shows the evolution with doping. Increasing carrier concentration diminishes both TMR and LMR (see Fig.\ref{FigDop}a)-d)). As in the case of thermal evolution, the LMR decreases faster than the TMR. At low doping,  the two configurations yield a similar amplitude. With increasing carrier concentration, the LMR becomes smaller than the TMR (see Fig.\ref{FigDop} b) and d)). 

Therefore, the unusual regime of the magnetoresistance detected by the present study emerges only at low temperature (when resistivity is dominated by its elastic component) and at low carrier concentration. Remarkably, even in this unusual context, the three components of the conductivity tensor keep the links expected by Eq.(1-3). This is demonstrated in the final panels of Fig.\ref{FigTemp} and Fig.\ref{FigDop}. The Hall mobility at a given magnetic field can be extracted using $\mu_H=\frac{1}{B}\frac{\sigma_{xy}}{\sigma_{\perp}}=\frac{1}{B}\frac{\rho_{xy}}{\rho_{\perp}}$ (see Fig.\ref{FigTemp}g)). The deduced $\mu_H(B)$ can then be compared with the field dependence of the longitudinal conductance $\sigma_{//}$ (see Fig.\ref{FigTemp}h)). As seen in the figure, there is a satisfactory agreement. This is the case of all samples at low doping levels, as shown in Fig.\ref{FigDop}e) and f). 

\begin{figure}[]
\begin{center}
\includegraphics[angle=0,width=8.5cm]{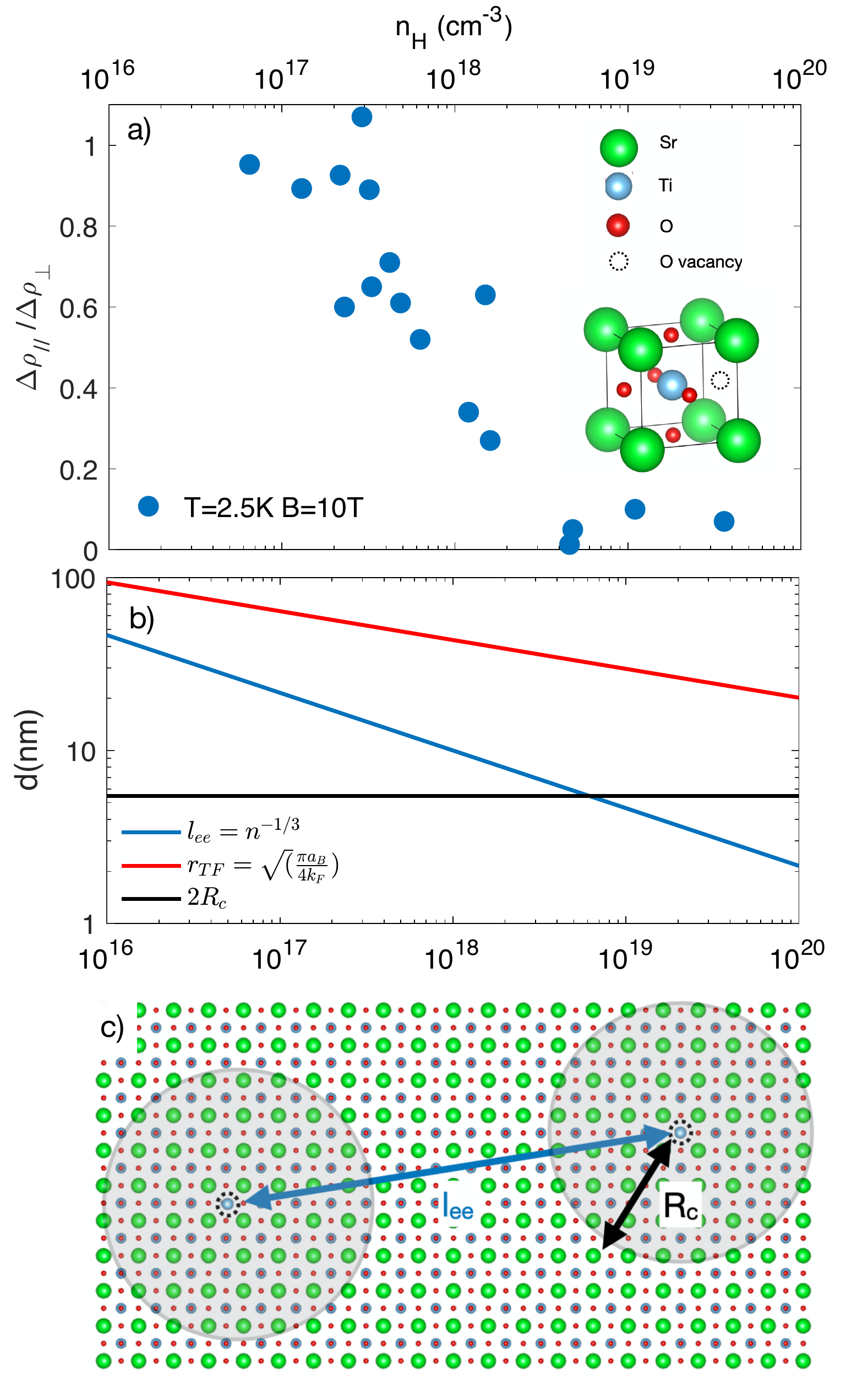}
\caption{ {\bf{Size of polar domains \textit{vs.} inter-electron distance in lightly doped SrTiO$_{3}$ }} : a) Doping dependence of the ratio of the LMR and TMR at B=10T for T=2.5K (in blue closed circles). Insert : sketch of the cubic unit cell of SrTiO$_{3}$ lattices in presence of an oxygen vacancy. b) Doping dependence of $\ell_{ee}$=$(n_H)^\frac{-1}{3}$ (the inter-electron distance), of the  screening length scale from charged impurities $r_{TF}=\sqrt{\frac{\pi a_B}{4 k_{F}}}$ compares with the polar domain diameter, 2R$_c$=5.4nm at low temperature. c) Sketch of the SrTiO$_{3}$ lattice in presence of two oxygen vacancies separated by a distance $\ell_{ee}$ and of the polar domains (in gray light) which form around each oxygen vacancies with a radius R$_c$. Below a critical doping where $\ell_{ee}$ becomes shorter than 2R$_c$  non zero LMR appears.} 
\label{Fig4}
\end{center}
\end{figure}

Thus, assuming that mobility smoothly evolves with magnetic field,  would explain both the quasi-linear non saturating TMR and the large finite LMR, which emerge at low doping. Fig.\ref{Fig4}a) shows the doping dependence of the LMR to TMR ratio ($\frac{\Delta \rho_{//}}{\Delta \rho_{\perp}}$ at $B=10$ T and $T=2.5$ K). Clearly, the finite LMR kicks in below a cut-off concentration and grows steadily with decreasing carrier density. The unusual magnetoresistance  of lightly doped SrTiO$_{3-\delta}$ is therefore restricted to carrier densities below a threshold of $3\times10^{18}$ cm$^{-3}$. As we will see below, a  clue to the origin of this phenomenon is provided by this boundary.

How can the mobility decrease  with magnetic field along both orientations? Why does this decrease happens in a restricted window of doping?  We will see below that a length scale specific to quantum paraelectrics plays a key role in finding answers to both of these questions.

A field-dependent mobility  has been previously invoked in other contexts \cite{Song2015, Fauque2018}. The time between collision events can become shorter in presence of magnetic field, because disorder is scanned differently at zero and finite magnetic fields. Compared to zero-field counterparts, charge carriers following a cyclotron orbit are more vulnerable to shallow scattering centers. Such a picture has been invoked to explain  the linear TMR in 3D high mobility dilute semiconductors \cite{Song2015} and the sub-quadratic TMR in semi-metals \cite{Fauque2018}. 

There are three already identified relevant length scales to the problem. These are, i) the Thomas-Fermi screening radius $r_{TF}=\sqrt{\frac{\pi a_B}{4 k_F}}$; ii) the magnetic length,  $\ell_{B}=\sqrt{\frac{\hbar}{eB}}$; and iii) the Fermi wavelength, $\lambda_F=2\pi k_F^{-1}$. When disorder is smooth and $r_{TF}$ is longer than the cyclotron radius ($r_{c}=\ell_{B}^2k_F$), the magnetic field, by quenching the kinetic energy of electrons in the plane of cyclotron motion, would guide them along the minimum of the electrostatic potential fluctuations \cite{Song2015}. This would lead to a decrease in mobility in the plane perpendicular to the magnetic field. The doping dependence of the Thomas-Fermi screening radius is shown in Fig.\ref{Fig4}b). Thanks to a Bohr radius as long as 600 nm in strontium titanate, $r_{TF}$ is remarkably long \cite{Behnia2015} and easily exceeds the cyclotron radius in a field of the order of Tesla. Therefore, shallow extended disorder, screened at zero-field will become visible as the cyclotron radius shrinks. One can invoke this picture to explain the quasi-linear TMR. However, the finite LMR and the low-field TMR remain both unexplained, because only the plane perpendicular of the orientation to the magnetic field is concerned.

In a polar crystal, defects, by distorting the lattice, generate electric dipoles. The typical length for correlation between such dipoles is set by $R_c$=$\frac{v_s}{\omega_{O}}$ (where $v_s$ and $\omega_{O}$ are the sound velocity and the frequency of the soft optical mode, respectively). In highly polarizable crystals, $\omega_{O}$ is small and $R_c$ can become remarkably long \cite{Vugmeister1990,Samara_2003}. In the specific case of strontium titanate  $v_s\simeq 7500$ m.s$^{-1}$ \cite{REHWALD1970607}, $\omega_{0}(300 K)\simeq 11$ meV and $\omega_0 (2$K$)\simeq 1.8$ meV \cite{Yamada1969}, therefore, $R_c$ varies from 0.5 nm at 300 K to 2.7 nm at 2 K. As a consequence, defects can cooperate with other defects over long distances to generate mesoscopic dipoles. In the case of a co-valent substitution, such as Sr$_{1-x}$Ca$_x$TiO$_{3}$, a Ca atom can break the local inversion symmetry. It can cooperate with other Ca sites within a range of $R_c$ to choose the same orientation for dipole alignment. When the Ca density exceeds a threshold, these domains percolate and generate a ferroelectric ground states. Remarkably, this critical density ($ x > 0.002$ \cite{Bednorz1984}) corresponds to  replacement of 1 out of 500 Sr atoms by Ca, that is when their average distance falls below $\frac{R_{c}}{a}$ (here, $a=0.39$ nm is the lattice parameter and  $\frac{R_{c}}{a} \approx 500^{-1/3}$).  In the case of  an oxygen vacancy, the donor, in addition to a local potential well, brings also a local dipole capable of cooperation with neighboring donors over long distances. 

Recent studies \cite{Rischau2017,wang2019charge} have confirmed the survival of dipolar physics in presence of dilute metallicity and the generation of ripples by electric dipoles inside the shallow Fermi sea. Specifically, it was found that in Sr$_{1-x}$Ca$_x$TiO$_{3-\delta}$, the ferroelectric-like alignment of dipoles is destroyed when there is more than one mobile electron per $7.9\pm0.6$ Ca atoms, the Fermi sea is dense enough to impede the percolation between polar domains. This threshold corresponds to an inter-electron distance approximately twice ($7.9^{-1/3} \approx 2 $) the inter-dipole distance. An oxygen vacancy (in addition to being a donor and an ionized point defect) generates an extended distortion of the size of $R_c$. This leads us to identify the origin of the doping window for the unusually isotropic magnetoresistance. If the inter-electron distance, ($\ell_{ee}$), which increases with decreasing carrier concentration, becomes significantly longer than $R_c$, then mobile carriers cannot adequately screen polar domains. Our data indicates that this is where the large quasi-isotropic magnetoresistance emerges. Fig.\ref{Fig4}b) shows the evolution of $\ell_{ee}=(n_H)^{-1/3}$ with doping. One can see that the threshold of $3\times10^{18}$ cm$^{-3}$ corresponds to $\ell_{ee}= 6.7$ nm. In other words, when the inter-electron distance becomes shorter than $2R_c$, the unusual magnetoresistance disappears, presumably because the Fermi sea is dense enough to impede the inhomogeneity generated by polar domains. This unusual MR is the largest at low temperature when resistivity is dominated by elastic scattering events. It vanishes with warming, when the inelastic $T^2$-term dominates over residual resistivity. This cross-over typically occurs around 15 K (see Fig.\ref{Fig1}b)). 

A possible solution to the mystery of the isotropic reduction of the mobility with magnetic field is offered by this length scale, which does not depend on the orientation of magnetic field. In presence of randomly oriented mesoscopic dipoles, the charge current does not align locally parallel to its macroscopic orientation. Instead, it will meander along a trajectory set by dipoles' electric field. The disorder affecting the whirling electrons will reduce mobility along different orientations. Remarkably, the inhomogeneity brought by these polar domains do not impede the existence of a percolated Fermi sea and the observation of quantum oscillations in this range of carrier concentration. A solid explanation of this apparent paradox remains a a task for future theoretical investigations. 

In summary, we found a large and quasi-isotropic magnetoresistance in lightly doped strontium titanate. We found that the longitudinal and the transverse magnetoresistance can be both explained in a picture where mobility changes with magnetic field and this arises as long the inter-electron distance is twice larger than the typical size of a polar domains. Other doped quantum paraelectrics, such as PbTe, KTaO$_{3}$ appear as potential candidates for displaying the same phenomenon. 
We thank M. Feigelman and B. Skinner for useful discussions. We acknowledge the support of the LNCMI-CNRS, member of the European Magnetic Field Laboratory (EMFL). This work was supported by JEIP-Coll\`{e}ge de France, by the Agence Nationale de la Recherche (ANR-18-CE92-0020-01; ANR-19-CE30-0014-04) and by a grant attributed by the Ile de France regional council. 

\bibliography{apssamp}

\end{document}